\documentclass[prx,twocolumn,showpacs]{revtex4}

\usepackage{graphicx}
\usepackage{dcolumn}
\usepackage{bm}
\usepackage{amsmath,amssymb}

\begin{document}


\title{Mott transition and electronic excitation of the Gutzwiller wavefunction}

\author{Masanori Kohno}
\email{KOHNO.Masanori@nims.go.jp}
\affiliation{International Center for Materials Nanoarchitectonics, National Institute for Materials Science, Tsukuba 305-0003, Japan}

\date{\today}

\begin{abstract}
The Mott transition is usually considered as resulting from the divergence of the effective mass of the quasiparticle in the Fermi-liquid theory; 
the dispersion relation around the Fermi level is considered to become flat toward the Mott transition. 
Here, to clarify the characterization of the Mott transition under the assumption of a Fermi-liquid-like ground state, 
the electron-addition excitation from the Gutzwiller wavefunction in the $t$-$J$ model is investigated on a chain, ladder, square lattice, and bilayer square lattice 
in the single-mode approximation using a Monte Carlo method. 
The numerical results demonstrate that an electronic mode that is continuously deformed from a noninteracting band at zero electron density 
loses its spectral weight and gradually disappears toward the Mott transition. 
It exhibits essentially the magnetic dispersion relation shifted by the Fermi momentum in the small-doping limit 
as indicated by recent studies for the Hubbard and $t$-$J$ models, 
even if the ground state is assumed to be a Fermi-liquid-like state exhibiting gradual disappearance of the quasiparticle weight. 
This implies that, rather than as the divergence of the effective mass or disappearance of the carrier density that is expected in conventional single-particle pictures, 
the Mott transition can be better understood as freezing of the charge degrees of freedom while the spin degrees of freedom remain active, 
even if the ground state is like a Fermi liquid. 
\end{abstract}


\maketitle
\section{Introduction} 
It is generally true that electrons in an interacting system are more difficult to move than those in a noninteracting system. 
As a result of the interaction, the effective mass increases, which implies that the dispersion relation around the Fermi level becomes flatter \cite{LandauFL,Nozieres,ImadaRMP}. 
The Mott transition is usually considered as an extreme case of this tendency: the electrons become immobile because of the effective-mass divergence. 
This picture, which is known as the Brinkman-Rice picture, was proposed in Ref. [\onlinecite{BrinkmanRice}], 
where the discontinuity of the momentum distribution function at the Fermi momentum (quasiparticle weight) was shown to decrease continuously to zero toward the Mott transition 
in the Gutzwiller approximation, which implies the divergence of the effective mass in the Fermi-liquid theory. 
\par
However, recent studies on electronic excitation near the Mott transition in the one-dimensional (1D), two-dimensional (2D), and ladder 
Hubbard and $t$-$J$ models \cite{KohnoRPP,Kohno1DHub,Kohno2DHub,Kohno1DtJ,Kohno2DtJ,KohnoDIS,KohnoAF,KohnoSpin,KohnoHubLadder} 
have indicated that an electronic mode in the Hubbard gap loses its spectral weight and exhibits the magnetic dispersion relation shifted by the Fermi momentum in the small-doping limit. 
This implies that the charge degrees of freedom freeze while the spin degrees of freedom remain active in the Mott transition. 
\par
Hence, the key question in this paper is how the electronic mode behaves if the ground state is like a Fermi liquid where the quasiparticle weight gradually disappears toward the Mott transition. 
In the Brinkman-Rice picture \cite{BrinkmanRice}, the gradual disappearance of the quasiparticle weight implies the gradual divergence of the effective mass, 
and flattening of the dispersion relation is expected. 
\par
In this paper, to resolve the above question, electron-addition excitation from the Gutzwiller wavefunction in the $t$-$J$ model is investigated on a chain, ladder, plane, and bilayer 
in the single-mode approximation using a Monte Carlo method. 
The numerical results demonstrate that an electronic mode that is continuously deformed from a noninteracting band at zero electron density 
gradually loses its spectral weight and exhibits essentially the momentum-shifted magnetic dispersion relation in the small-doping limit, 
even if the ground state is assumed to be a Fermi-liquid-like state that exhibits gradual disappearance of the quasiparticle weight toward the Mott transition. 
\par
This suggests that this characteristic of the Mott transition \cite{KohnoRPP,Kohno1DHub,Kohno2DHub,Kohno1DtJ,Kohno2DtJ,KohnoDIS,KohnoAF,KohnoSpin,KohnoHubLadder} 
is not highly sensitive to the ground-state properties, but would be general and fundamental in the Mott transition. 
Thus, the Mott transition can be better understood in terms of this characteristic 
\cite{KohnoRPP,Kohno1DHub,Kohno2DHub,Kohno1DtJ,Kohno2DtJ,KohnoDIS,KohnoAF,KohnoSpin,KohnoHubLadder}, 
rather than conventional single-particle pictures, such as the divergence of the effective mass or disappearance of the carrier density \cite{ImadaRMP}, 
regardless of whether the ground state is like a Fermi liquid or not. 
\section{Model and method} 
\subsection{Model and parameters} 
The $t$-$J$ model is defined by the following Hamiltonian:
\begin{eqnarray}
{\cal H}&=&-\sum_{\langle i,j\rangle,\sigma}t_{{i,j}}\left({\tilde c}^{\dagger}_{i,\sigma}{\tilde c}_{j,\sigma}+{\mbox {H.c.}}\right)\nonumber\\
&&+\sum_{\langle i,j\rangle}J_{i,j}\left({\bm S}_i\cdot{\bm S}_j-\frac{1}{4}n_in_j\right)
-\mu\sum_{i,\sigma}n_{i,\sigma},
\end{eqnarray}
where ${\tilde c}_{i,\sigma}$ denotes the annihilation operator of an electron with spin $\sigma$ at site $i$ 
under the constraint of no double occupancy, and $\langle i,j\rangle$ means that sites $i$ and $j$ are nearest neighbors. 
Here, $n_{i,\sigma}$ and ${\bm S}_i$ denote the number operator with spin $\sigma$ and the spin operator at site $i$, respectively, 
and $n_i=\sum_{\sigma}n_{i,\sigma}$. 
In this paper, we consider the $t$-$J$ models on a chain ($t_{i,j}=t$, $J_{i,j}=J$), planar square lattice ($t_{i,j}=t$, $J_{i,j}=J$), 
ladder ($t_{i,j}=t$ and $J_{i,j}=J$ in the legs; $t_{i,j}=t_{\perp}$ and $J_{i,j}=J_{\perp}$ in the rungs), and 
bilayer square lattice ($t_{i,j}=t$ and $J_{i,j}=J$ in the layers; $t_{i,j}=t_{\perp}$ and $J_{i,j}=J_{\perp}$ between the layers). 
\par
Hereafter, the numbers of sites and electrons are denoted by $N_{\rm s}$ and $N_{\rm e}$, respectively. 
The electron density and doping concentration are defined as $n=N_{\rm e}/N_{\rm s}$ and $\delta=1-n$, respectively. 
At half filling, $n=1$ and $\delta=0$. 
For a ladder and bilayer, the momentum in the interchain or interlayer direction is denoted by $k_{\perp}$. 
The momenta on a ladder and bilayer are represented as $(k_x,k_{\perp})$ and $(k_x,k_y,k_{\perp})$, respectively. 
The shorthand notations ${\bm 0}$ and ${\bm \pi}$ are used for $(0,0)$ and $(\pi,\pi)$, respectively. 
As a compact notation, $k_x$ and $k_y$ are sometimes denoted by $k_1$ and $k_2$, respectively. 
\par
In this paper, the numerical results for $J/t=0.5$ on a chain and plane; $J/t=0.25$, $t_{\perp}/t=2$, and $J_{\perp}/t=1$ on a ladder; 
and $J/t=0.25$, $t_{\perp}/t=4$, and $J_{\perp}/t=4$ on a bilayer with $t>0$ are presented. 
The calculations were performed under periodic boundary conditions 
on clusters of $N_{\rm s}=120$ for the chain and ladder, $N_{\rm s}=400$ for the plane, and $N_{\rm s}=200$ for the bilayer. 
Typically, several millions of samples were generated following several hundreds of sweeps in the Monte Carlo calculations. 
\subsection{Gutzwiller wavefunction} 
In this paper, the ground state is assumed to be the Gutzwiller wavefunction $|\Phi\rangle$, defined as \cite{GutzwillerWF} 
\begin{equation}
|\Phi\rangle={\rm P}_{\rm d}|{\rm FS}\rangle,\quad
|{\rm FS}\rangle=\prod_{\sigma}\prod_{{\bm k}\in\mbox{Fermi sea}}c^{\dagger}_{\bm k,\sigma}|0\rangle,
\end{equation}
where $c^{\dagger}_{{\bm k},\sigma}$ denotes the creation operator of an electron with momentum ${\bm k}$ and spin $\sigma$, and $|0\rangle$ represents the vacuum. 
Here, ${\rm P}_{\rm d}$ denotes the projection operator that forbids double occupancy. 
The excitation energy $\varepsilon({\bm k})$ and spectral weight $W({\bm k})$ 
of the electron-addition excited state ${\tilde c}^{\dagger}_{{\bm k},\sigma}|\Phi\rangle$ averaged with respect to spin are obtained as follows: 
\begin{eqnarray}
\label{eq:ek}
\varepsilon({\bm k})&=&\frac{1}{2}\sum_{\sigma}\frac{\langle\Phi|{\tilde c}_{{\bm k},\sigma}{\cal H}{\tilde c}^{\dagger}_{{\bm k},\sigma}|\Phi\rangle}
{\langle\Phi|{\tilde c}_{{\bm k},\sigma}{\tilde c}^{\dagger}_{{\bm k},\sigma}|\Phi\rangle}
-\frac{\langle\Phi|{\cal H}|\Phi\rangle}{\langle\Phi|\Phi\rangle},\\
\label{eq:Wk}
W({\bm k})&=&\frac{1}{2}\sum_{\sigma}\frac{\langle\Phi|{\tilde c}_{{\bm k},\sigma}{\tilde c}^{\dagger}_{{\bm k},\sigma}|\Phi\rangle}{\langle\Phi|\Phi\rangle},
\end{eqnarray}
where ${\tilde c}_{{\bm k},\sigma}$ denotes $c_{{\bm k},\sigma}$ with the constraint of no double occupancy. 
The expectation value of an operator ${\cal O}$ by $|\Phi\rangle$ can be evaluated as the sample average of weight $w_i$ 
for configuration $|i\rangle$ generated with probability $p_i$ using a Monte Carlo method \cite{Ceperley,GrosVMC}: 
\begin{equation}
\frac{\langle\Phi|{\cal O}|\Phi\rangle}{\langle\Phi|\Phi\rangle}=\sum_iw_ip_i, 
\end{equation}
where 
\begin{eqnarray}
w_i&=&\sum_j\langle j|{\cal O}|i\rangle\frac{\langle\Phi|j\rangle}{\langle\Phi|i\rangle},\\
p_i&=&\frac{|\langle i|\Phi\rangle|^2}{\sum_l|\langle l|\Phi\rangle|^2}. 
\end{eqnarray}
\par
It should be noted that ${\tilde c}^{\dagger}_{{\bm k},\sigma}$ and ${\rm P_d}$ commute: 
\begin{equation}
\label{eq:projection}
{\tilde c}^{\dagger}_{{\bm k},\sigma}{\rm P_d}|{\rm FS}\rangle={\rm P_d}c^{\dagger}_{{\bm k},\sigma}|{\rm FS}\rangle, 
\end{equation}
because ${\tilde c}^{\dagger}_{i,\sigma}{\rm P}_{\rm d}|\alpha\rangle_i={\rm P}_{\rm d}c^{\dagger}_{i,\sigma}|\alpha\rangle_i$, 
where $c^{\dagger}_{i,\sigma}$ and $|\alpha\rangle_i$ denote the creation operator of an electron with spin $\sigma$ and 
a state (0, $\uparrow$, $\downarrow$, or $\uparrow\downarrow$) at site $i$, respectively. 
Thus, the first term on the right-hand side of Eq. (\ref{eq:ek}) can be calculated as the energy of the Gutzwiller wavefunction 
where an electron with momentum ${\bm k}$ and spin $\sigma$ is added to the Fermi sea prior to projection. 
This can significantly reduce the computational complexity of the electron-addition energy. 
On the other hand, ${\tilde c}_{{\bm k},\sigma}{\rm P_d}|{\rm FS}\rangle\ne{\rm P_d}c_{{\bm k},\sigma}|{\rm FS}\rangle$ in general. 
\par
The chemical potential $\mu$ at $N_{\rm e}=m$ can be calculated as follows: 
\begin{equation}
\mu=(E_{m+1}-E_{m-1})/2, 
\end{equation}
where $E_{m\pm 1}$ denotes the ground-state energy at $N_{\rm e}=m\pm 1$ for $\mu=0$. 
In this paper, because the ground state is assumed to be expressed as the Gutzwiller wavefunction, 
$\mu$ can be calculated using the lowest energies of the Gutzwiller wavefunction 
with an electron added to ($E_{m+1}$) and removed from ($E_{m-1}$) the Fermi sea at $N_{\rm e}=m$ prior to projection. 
Similarly, the Fermi momentum ${\bm k}_{\rm F}$ can be determined as the momentum 
where the dispersion relation of the Gutzwiller wavefunction with an electron added or removed prior to projection crosses the Fermi level. 
\subsection{Single-mode approximation for spectral function} 
\begin{figure*}
\includegraphics[width=\linewidth]{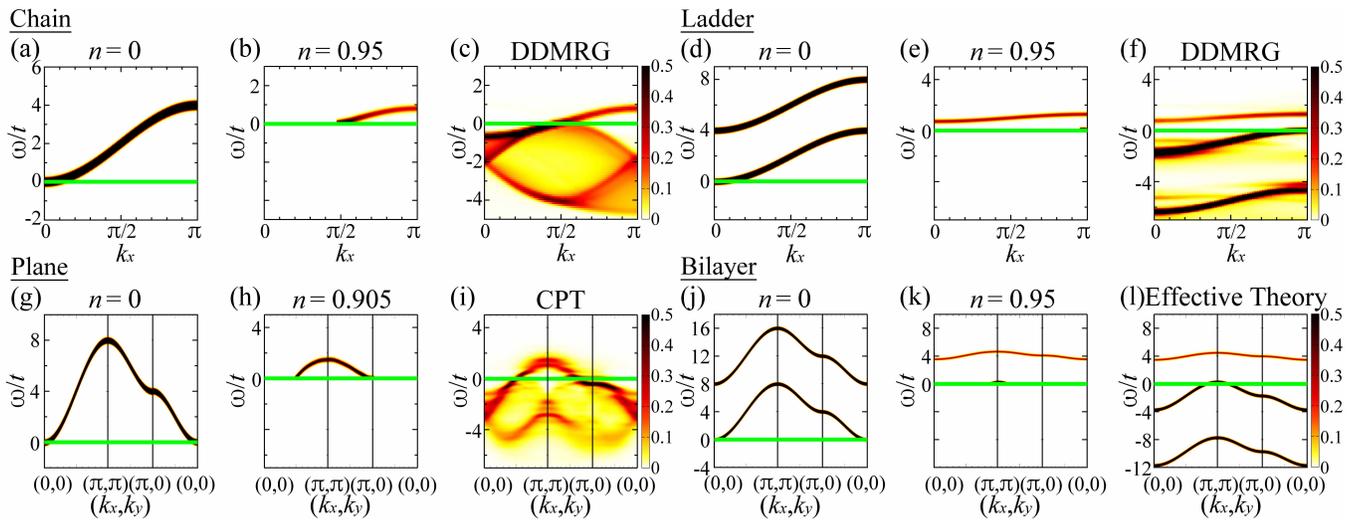}
\caption{Spectral function on chain [(a)--(c)], ladder at $k_{\perp}=0, \pi$ [(d)--(f)], plane [(g)--(i)], and bilayer at $k_{\perp}=0, \pi$ [(j)--(l)]. 
(a), (d), (g), (j) $A({\bm k},\omega)t$ at $n=0$ on chain [(a)], ladder [(d)], plane [(g)], and bilayer [(j)] [Eqs. (\ref{eq:ek0chain}) and (\ref{eq:ek0ladder})]. 
(b), (e), (h), (k) $A^{\rm s}({\bm k},\omega)t$ for $\omega>0$ on chain at $n=0.95$ [(b)], ladder at $n=0.95$ [(e)], plane at $n=0.905$ [(h)], and bilayer at $n=0.95$ [(k)], 
where the dispersion relation and spectral weight by cubic spline interpolation in Fig. \ref{fig:EkWk} are used. 
(c), (f) $A({\bm k},\omega)t$ on chain at $n=0.95$ [(c)] and ladder at $n=0.95$ [(f)] 
obtained using the non-Abelian DDMRG method 
with 240 density-matrix eigenstates on a 120-site cluster \cite{Kohno1DtJ,KohnoDIS}. 
(i) $A({\bm k},\omega)t$ at $n=0.905$ on plane obtained using CPT with $4\times 4$-site clusters \cite{Kohno2DtJ}. 
(l) $A({\bm k},\omega)t$ at $n=0.95$ on bilayer in the effective theory near half filling for $t_{\perp}\gg t$ and $J_{\perp}\gg J$ 
[Eqs. (\ref{eq:ektlimit}) and (\ref{eq:ektladderlimit})] \cite{KohnoDIS}. 
The green lines represent the Fermi level ($\omega=0$). Gaussian broadening with a standard deviation of $0.1t$ is used.}
\label{fig:Akw}
\end{figure*}
The spectral function is defined as 
\begin{eqnarray}
\label{eq:Akw}
A({\bm k},\omega)&=&\frac{1}{2}\sum_{\sigma,l}\frac{|\langle l|{\tilde c}^{\dagger}_{{\bm k},\sigma}|{\rm GS}\rangle|^2}{\langle{\rm GS}|{\rm GS}\rangle}\delta(\omega-\varepsilon_l)\nonumber\\
&+&\frac{1}{2}\sum_{\sigma,l}\frac{|\langle l|{\tilde c}_{{\bm k},\sigma}|{\rm GS}\rangle|^2}{\langle{\rm GS}|{\rm GS}\rangle}\delta(\omega+\varepsilon_l), 
\end{eqnarray}
where $\varepsilon_l$ denotes the excitation energy of the normalized eigenstate $|l\rangle$ from the ground state $|{\rm GS}\rangle$. 
In this paper, the single-mode approximation is employed, where the electron-addition spectral function [$A({\bm k},\omega)$ for $\omega>0$] is approximated as 
\begin{equation}
\label{eq:AkwSM}
A^{\rm s}({\bm k},\omega)=W({\bm k})\delta(\omega-\varepsilon({\bm k})). 
\end{equation}
If the excitation is essentially represented by a dominant mode, 
the single-mode approximation can capture the essential excitation feature. 
It has been shown that the electron-addition excitation ($\omega>0$) can be effectively represented by a single mode 
in the $t$-$J$ models near half filling (at each $k_{\perp}$ for the ladder and bilayer) [Figs. \ref{fig:Akw}(c), \ref{fig:Akw}(f), \ref{fig:Akw}(i), and \ref{fig:Akw}(l)] \cite{Kohno1DtJ, Kohno2DtJ, KohnoDIS}. 
However, if the spectral weight is spread over a wide range of $\omega$ at each ${\bm k}$, 
as observed in the electron-removal excitation ($\omega<0$) in the $t$-$J$ and Hubbard models [Figs. \ref{fig:Akw}(c), \ref{fig:Akw}(f), and \ref{fig:Akw}(i)] 
\cite{KohnoRPP, Kohno1DHub, Kohno2DHub, Kohno1DtJ, Kohno2DtJ, KohnoDIS, KohnoHubLadder, KohnoAF, KohnoSpin}, 
the single-mode approximation exhibits a single peak at the weighted mean value of $\omega$. 
The mode for $\omega<0$ in this approximation can exhibit an excitation gap even if the true excitation is gapless. 
Thus, in this paper, we only consider the electron-addition excitation ($\omega>0$), which exhibits a significant characteristic toward the Mott transition. 
\section{Results and discussions} 
\subsection{Spectral function} 
\label{sec:spectralfunction}
At zero electron density ($n=0$), the spectral function for an added electron is the same as that in a noninteracting system 
[Figs. \ref{fig:Akw}(a), \ref{fig:Akw}(d), \ref{fig:Akw}(g), and \ref{fig:Akw}(j)], because no other electrons exist. 
As the electron density increases (i.e., the chemical potential is increased), the Fermi level moves into the (lower) band 
and the spectral function in the single-mode approximation [$A^{\rm s}({\bm k},\omega)$; Eq. (\ref{eq:AkwSM})] for $\omega>0$ 
becomes that indicated in Figs. \ref{fig:Akw}(b), \ref{fig:Akw}(e), \ref{fig:Akw}(h), and \ref{fig:Akw}(k). 
\par
The validity of the results is confirmed by their comparison with the results obtained using the non-Abelian dynamical density-matrix renormalization-group (DDMRG) method 
for the chain [Fig. \ref{fig:Akw}(c)] \cite{Kohno1DtJ} and ladder [Fig. \ref{fig:Akw}(f)] \cite{KohnoDIS}, those obtained using the cluster perturbation theory (CPT) 
for the plane [Fig. \ref{fig:Akw}(i)] \cite{Kohno2DtJ}, and those of the effective theory near half filling for $t_{\perp}\gg t$ and $J_{\perp}\gg J$ 
for the bilayer [Fig. \ref{fig:Akw}(l)] \cite{KohnoDIS}. 
In the effective theory \cite{KohnoDIS}, the dispersion relation at $k_{\perp}=\pi$ for $\omega>0$ is obtained as follows: 
\begin{equation}
\label{eq:ektlimit}
\omega=-J\sum_{i=1}^d \cos k_i+J_{\perp},
\end{equation}
and that of the other modes can be expressed as 
\begin{equation}
\label{eq:ektladderlimit}
\omega=-t\sum_{i=1}^d(\cos k_i-\cos k_{{\rm F}i})+t_{\perp}(\cos k_{\perp}-1)
\end{equation}
on the ladder ($d=1$) and bilayer ($d=2$), where the $x$ and $y$ components of ${\bm k}_{\rm F}$ are denoted 
by $k_{{\rm F}1}$ and $k_{{\rm F}2}$, respectively. 
In Fig. \ref{fig:Akw}(l), the spectral weight at each ${\bm k}$ is approximated as $1.5\delta$ for $\omega>0$ at $k_{\perp}=\pi$, 
$0.5-\delta$ for $\omega<0$ at $k_{\perp}=\pi$, and 0.5 at $k_{\perp}=0$. The contributions from the continua (multiparticle processes) are neglected. 
\par
In the following sections, the changes in the dispersion relation and spectral weight with the electron density are discussed. 
\subsection{Dispersion relation} 
\label{sec:dispersion}
\begin{figure*}
\includegraphics[width=\linewidth]{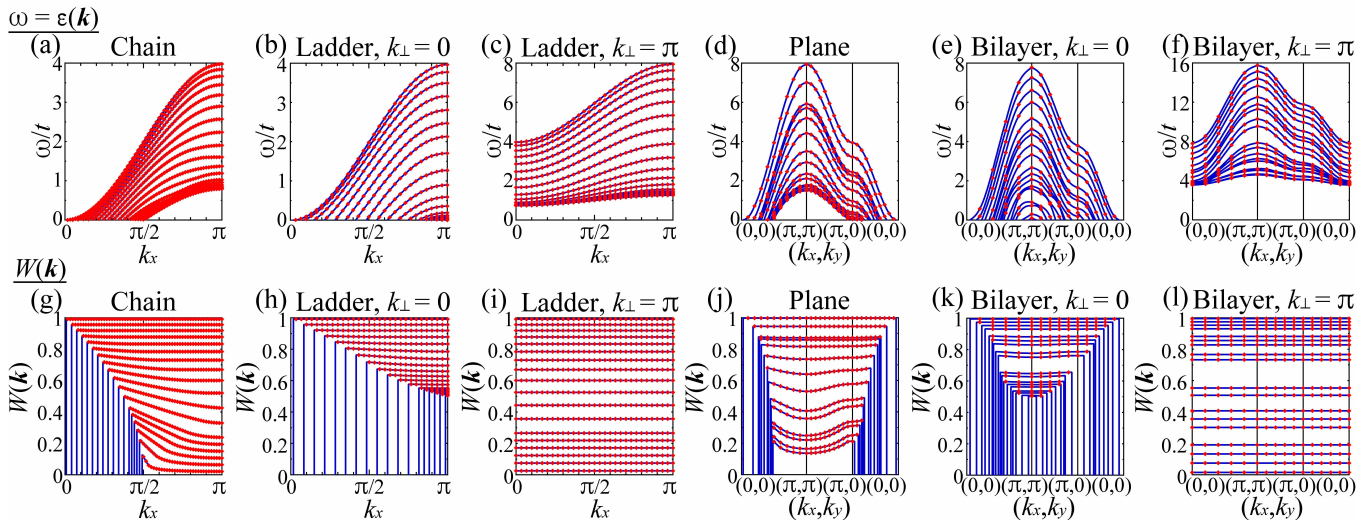}
\caption{Dispersion relation $\omega=\varepsilon({\bm k})$ [(a)--(f)] and spectral weight $W({\bm k})$ [(g)--(l)] for $\omega>0$ on chain [(a), (g)], 
ladder at $k_{\perp}=0$ [(b), (h)] and $\pi$ [(c), (i)], plane [(d), (j)], and bilayer at $k_{\perp}=0$ [(e), (k)] and $\pi$ [(f), (l)]. 
The red diamonds denote the Monte Carlo results. The blue lines indicate the cubic spline interpolation. 
For the chain and ladder [(a)--(c), (g)--(i)], $n\approx$ 0.017, 0.083, 0.150, 0.217, 0.283, 0.350, 0.417, 0.483, 0.550, 0.617, 
0.683, 0.750, 0.817, 0.850, 0.883, 0.917, 0.950, and 0.983 from above. 
For the plane [(d), (j)], $n=$0.005, 0.105, 0.225, 0.245, 0.305, 0.405, 0.505, 0.605, 0.705, 0.745, 0.825, 0.845, 0.885, and 0.905 from above. 
For the bilayer [(e), (f), (k), (l)], $n=$0.01, 0.05, 0.09, 0.13, 0.21, 0.25, 0.29, 0.37, 0.41, 0.59, 0.63, 0.71, 0.75, 0.79, 0.87, 0.91, 0.95, and 0.99 from above.}
\label{fig:EkWk}
\end{figure*}
\begin{figure}
\includegraphics[width=\linewidth]{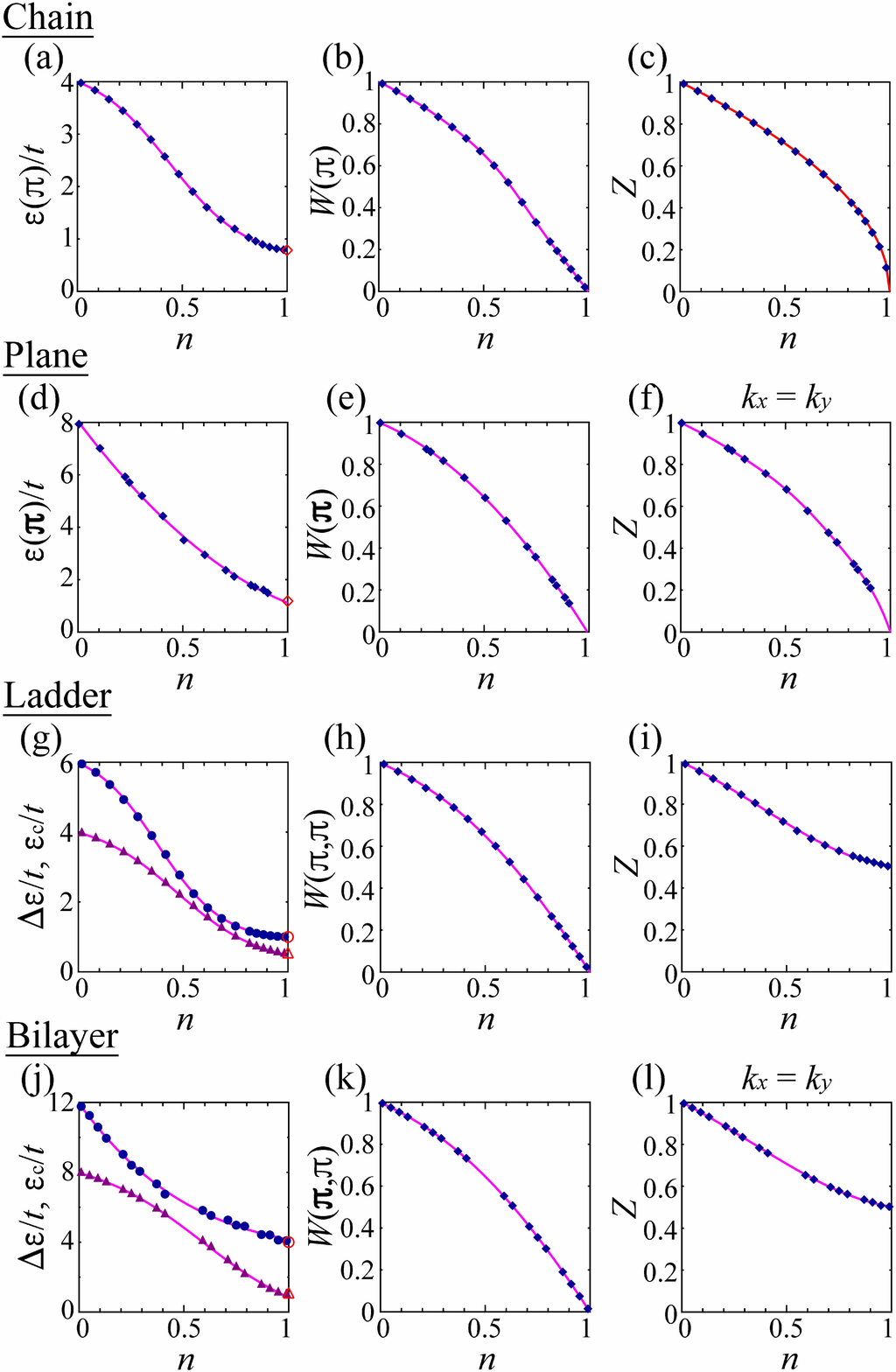}
\caption{Characteristic energies and spectral weights as a function of electron density. 
(a) $\varepsilon(\pi)/t$, (b) $W(\pi)$, and (c) $Z$ on chain. 
(d) $\varepsilon({\bm \pi})/t$, (e) $W({\bm \pi})$, and (f) $Z$ for $k_x=k_y$ on plane. 
(g) $\varepsilon_{\rm c}/t$ (solid blue circles) and $\Delta \varepsilon/t$ (solid purple triangles), 
(h) $W(\pi,\pi)$, and (i) $Z$ on ladder. 
(j) $\varepsilon_{\rm c}/t$ (solid blue circles) and $\Delta \varepsilon/t$ (solid purple triangles), 
(k) $W({\bm \pi}, \pi)$, and (l) $Z$ for $k_x=k_y$ on bilayer. 
The solid blue diamonds, solid blue circles, and solid purple triangles indicate the Monte Carlo results. 
The red curve in (c) represents the exact result for the Gutzwiller wavefunction on the chain \cite{MetznerPRL,MetznerPRB}. 
The magenta curves are guides for the eye. 
In (a) and (d), the open red diamonds at $n=1$ indicate $e_{\rm 1D}(\pi/2)/t$ with $v_{\rm 1D}=\pi J/2$ [\onlinecite{desCloizeaux}] [(a)] 
and $e_{\rm 2D}({\bm \pi}/2)/t$ with $v_{\rm 2D}=1.18\sqrt{2}J$ [\onlinecite{Singh}] [(d)]. 
In (g) and (j), the open red circles at $n=1$ indicate $e_{\rm c}/t$ on the ladder [(g)] and bilayer [(j)], 
and the open red triangles at $n=1$ indicate $\Delta e/t$ on the ladder [(g)] and bilayer [(j)].}
\label{fig:ZWe}
\end{figure}
At $n=0$, because the added electron behaves as a noninteracting electron, the dispersion relation can be expressed as 
\begin{equation}
\label{eq:ek0chain}
\omega=-2t\sum_{i=1}^d(\cos k_i-1)
\end{equation}
on the chain ($d=1$) and plane ($d=2$), and as 
\begin{equation}
\label{eq:ek0ladder}
\omega=-2t\sum_{i=1}^d(\cos k_i-1)-t_{\perp}(\cos k_{\perp}-1)
\end{equation}
on the ladder ($d=1$) and bilayer ($d=2$), where the Fermi level is set to the bottom of the (lower) band 
[Figs. \ref{fig:Akw}(a), \ref{fig:Akw}(d), \ref{fig:Akw}(g), and \ref{fig:Akw}(j)]. 
\par
As illustrated in Figs. \ref{fig:EkWk}(a)--\ref{fig:EkWk}(f), the dispersion relation of the electron-addition excitation [$\omega=\varepsilon({\bm k})$] 
changes continuously as the electron density increases from $n=0$ [Eqs. (\ref{eq:ek0chain}) and (\ref{eq:ek0ladder})]. 
To clarify the electron-density dependence, Figs. \ref{fig:ZWe}(a), \ref{fig:ZWe}(d), \ref{fig:ZWe}(g), and \ref{fig:ZWe}(j) 
display the characteristic energies: $\varepsilon(\pi)$ on the chain, $\varepsilon({\bm \pi})$ on the plane, 
$\Delta\varepsilon$ and $\varepsilon_{\rm c}$ on the ladder and bilayer. 
Here, $\Delta\varepsilon$ and $\varepsilon_{\rm c}$ denote the bandwidth and band center, respectively, which are defined as 
\begin{eqnarray}
\label{eq:bandwidth}
\Delta\varepsilon&=&\varepsilon({\bm Q}_{\rm max},\pi)-\varepsilon({\bm Q}_{\rm min},\pi),\\
\label{eq:bandcenter}
\varepsilon_{\rm c}&=&[\varepsilon({\bm Q}_{\rm max},\pi)+\varepsilon({\bm Q}_{\rm min},\pi)]/2,
\end{eqnarray}
where $ {\bm Q}_{\rm min}$ and ${\bm Q}_{\rm max}$ represent $0$ and $\pi$ on the ladder, and ${\bm 0}$ and ${\bm \pi}$ on the bilayer. 
\par
For the chain and plane, the dispersion relation continues to disperse even in the limit of $n\rightarrow 1$ 
[Figs. \ref{fig:EkWk}(a), \ref{fig:EkWk}(d), \ref{fig:ZWe}(a), and \ref{fig:ZWe}(d)]. 
This implies that the mode for $\omega>0$, which is continuously deformed from that of a noninteracting electron at $n=0$, 
does not become flat toward the Mott transition, in contrast to the conventional single-particle picture of the effective-mass divergence. 
\par
For the ladder and bilayer, the dispersion relation at $k_{\perp}=0$ shrinks to $\omega\rightarrow 0$ 
at $k_x=\pi$ and $(k_x,k_y)={\bm \pi}$, respectively, in the limit of $n\rightarrow 1$ [Figs. \ref{fig:EkWk}(b) and \ref{fig:EkWk}(e)], 
whereas the dispersion relation at $k_{\perp}=\pi$ continues to disperse [Figs. \ref{fig:EkWk}(c), \ref{fig:EkWk}(f), \ref{fig:ZWe}(g), and \ref{fig:ZWe}(j)]. 
Although these features are similar to those of the transition from a metal to a band insulator, 
the spectral weight at $k_{\perp}=\pi$ gradually disappears toward the Mott transition, in contrast to the conventional band picture, 
as shown in Sec. \ref{sec:spectralweight}. 
\par
To clarify the nature of the electron-addition excitation in the limit of $n\rightarrow 1$, we consider the excitation at half filling ($n=1$), 
where the $t$-$J$ model is reduced to the Heisenberg model. 
For the chain and plane, the dominant spin excitation exhibits the following spin-wave dispersion relation \cite{desCloizeaux,AndersonSW}: 
\begin{equation}
\label{eq:eskchain}
e_{\rm 1D}(k_x)=v_{\rm 1D}|\sin k_x|
\end{equation}
on the chain and 
\begin{equation}
\label{eq:eskplane}
e_{\rm 2D}(k,k)=\sqrt{2}v_{\rm 2D}|\sin k|
\end{equation}
for $k_x=k_y=k$ on the square lattice, 
where the spin-wave velocities of the Heisenberg models on the chain and square lattice have been obtained as 
$v_{\rm 1D}=\pi J/2$ [\onlinecite{desCloizeaux}] and $v_{\rm 2D}=1.18(2)\sqrt 2 J$ [\onlinecite{Singh}], respectively. 
As illustrated in Figs. \ref{fig:ZWe}(a) and \ref{fig:ZWe}(d), $\varepsilon(\pi)$ and $\varepsilon({\bm \pi})$ in the limit of $n\rightarrow 1$ 
reasonably well approach $e_{\rm 1D}(\pi/2)$ and $e_{\rm 2D}({\bm \pi}/2)$ (open red diamonds) on the chain and plane, respectively. 
\par
For the ladder and bilayer, the dispersion relation of the spin excitation at $k_{\perp}=\pi$ for $J_{\perp}\gg J$ can effectively be expressed as 
\begin{equation}
\label{eq:eskladder}
e_{\rm eff}({\bm k})=J\sum_{i=1}^d\cos k_i+J_{\perp}
\end{equation}
on the ladder ($d=1$) and bilayer ($d=2$) \cite{KohnoDIS}. 
As indicated in Figs. \ref{fig:ZWe}(g) and \ref{fig:ZWe}(j), 
$\Delta\varepsilon$ (solid purple triangles) and $\varepsilon_{\rm c}$ (solid blue circles) in the limit of $n\rightarrow 1$ are reduced to 
the bandwidth $\Delta e$ (open red triangles) and band center $e_{\rm c}$ (open red circles) of the spin excitation at half filling, respectively, which are defined as 
\begin{eqnarray}
\label{eq:bandwidths}
\Delta e&=&e_{\rm eff}({\bm Q}_{\rm min},\pi)-e_{\rm eff}({\bm Q}_{\rm max},\pi),\\
\label{eq:bandcenters}
e_{\rm c}&=&[e_{\rm eff}({\bm Q}_{\rm min},\pi)+e_{\rm eff}({\bm Q}_{\rm max},\pi)]/2. 
\end{eqnarray}
\par
The above results are consistent with the general relationship between the dispersion relation of the spin excitation in a Mott insulator 
($N_{\rm e}=N_{\rm s}$) and that of the electron-addition excitation in the small-doping limit ($N_{\rm e}=N_{\rm s}-1$) shown in Ref. [\onlinecite{KohnoDIS}]: 
spin-excited states in a Mott insulator can emerge in the electron-addition spectrum outside the Fermi surface, 
exhibiting the magnetic dispersion relation shifted by ${\bm k}_{\rm F}$ in the small-doping limit 
\cite{KohnoRPP,Kohno1DHub,Kohno2DHub,Kohno1DtJ,Kohno2DtJ,KohnoAF,KohnoSpin,KohnoDIS,KohnoHubLadder}. 
By applying this relationship, the dispersion relation of the electron-addition excitation in the small-doping limit is expected to be 
\begin{equation}
\label{eq:ekchainlimit}
\omega=-v_{\rm 1D}\cos k_x
\end{equation}
for $\pi/2<k_x<3\pi/2$ on the chain [${\bm k}_{\rm F}=\pi/2$; Eq. (\ref{eq:eskchain})], 
\begin{equation}
\label{eq:ekplanelimit}
\omega=-\sqrt{2}v_{\rm 2D}\cos k
\end{equation}
for $\pi/2<k<3\pi/2$ along $k_x=k_y=k$ on the plane [${\bm k}_{\rm F}={\bm \pi}/2$; Eq. (\ref{eq:eskplane})], 
and Eq. (\ref{eq:ektlimit}) on the ladder at $k_{\perp}=\pi$ [$d=1$; ${\bm k}_{\rm F}=(\pi,0)$; Eq. (\ref{eq:eskladder})] 
and bilayer at $k_{\perp}=\pi$ [$d=2$; ${\bm k}_{\rm F}=({\bm \pi},0)$; Eq. (\ref{eq:eskladder})]. 
The results that are obtained simply by assuming that the ground state is the Gutzwiller wavefunction 
[Figs. \ref{fig:Akw}(b), \ref{fig:Akw}(e), \ref{fig:Akw}(h), \ref{fig:Akw}(k), \ref{fig:EkWk}(a), \ref{fig:EkWk}(c), \ref{fig:EkWk}(d), \ref{fig:EkWk}(f), 
\ref{fig:ZWe}(a), \ref{fig:ZWe}(d), \ref{fig:ZWe}(g), and \ref{fig:ZWe}(j)] 
agree reasonably well with this behavior [Eqs. (\ref{eq:ektlimit}), (\ref{eq:ekchainlimit}), and (\ref{eq:ekplanelimit}); Fig. \ref{fig:Akw}(l)]. 
\subsection{Spectral weight} 
\label{sec:spectralweight}
At $n=0$, $W({\bm k})=1$ because the electron-addition excitation is the same as that of a noninteracting system 
[Figs. \ref{fig:Akw}(a), \ref{fig:Akw}(d), \ref{fig:Akw}(g), and \ref{fig:Akw}(j)]. 
As the electron density increases, the spectral weight outside the Fermi surface gradually decreases, as illustrated in Figs. \ref{fig:EkWk}(g)--\ref{fig:EkWk}(l). 
To clarify the electron-density dependence, Figs. \ref{fig:ZWe}(b), \ref{fig:ZWe}(e), \ref{fig:ZWe}(h), and \ref{fig:ZWe}(k) display the characteristic spectral weights: 
$W(\pi)$ on the chain, $W({\bm \pi})$ on the plane, $W(\pi,\pi)$ on the ladder, and $W({\bm \pi},\pi)$ on the bilayer. 
At $n=1$, $W({\bm k})=0$ because an electron cannot be added to the ground state with the constraint of no double occupancy. 
The Hubbard gap can be regarded as infinitely large. 
\par
The spectral weights at $k_{\perp}=0$ on the ladder and bilayer remain nonzero even in the limit of $n\rightarrow 1$ [Figs. \ref{fig:EkWk}(h) and \ref{fig:EkWk}(k)], 
as in the case of the transition from a metal to a band insulator. 
However, the spectral weights on the chain, plane, and ladder at $k_{\perp}=\pi$, as well as on the bilayer at $k_{\perp}=\pi$ 
gradually disappear toward the Mott transition ($n\rightarrow 1$) 
[Figs. \ref{fig:EkWk}(g), \ref{fig:EkWk}(i), \ref{fig:EkWk}(j), \ref{fig:EkWk}(l), \ref{fig:ZWe}(b), \ref{fig:ZWe}(e), \ref{fig:ZWe}(h), and \ref{fig:ZWe}(k)]. 
\par
These results imply the following: 
For the chain and plane, the dispersing mode crossing the Fermi level (Sec. \ref{sec:dispersion}), 
which is continuously deformed from that of a noninteracting electron at $n=0$, 
loses its spectral weight and gradually disappears toward the Mott transition 
without flattening of the dispersion relation \cite{KohnoRPP,Kohno1DHub,Kohno2DHub,Kohno1DtJ,Kohno2DtJ,KohnoAF,KohnoSpin}. 
For the ladder and bilayer, the mode at $k_{\perp}=\pi$, which is continuously deformed from the noninteracting antibonding band ($k_{\perp}=\pi$) at $n=0$, 
persists as a dispersing mode in the metallic phase (Sec. \ref{sec:dispersion}) but loses its spectral weight and gradually disappears 
as $n\rightarrow 1$ \cite{KohnoDIS,KohnoHubLadder}, 
contrary to the conventional band picture in which the number of bands is considered to be determined by the number of atomic orbitals in a unit cell \cite{AshcroftMermin} 
and invariant with the electron density provided that symmetry breaking does not occur (neither emergence nor disappearance of a band is expected). 
\subsection{Quasiparticle weight} 
The momentum distribution function is defined as 
\begin{equation}
n({\bm k})=\frac{1}{2}\sum_{\sigma}\frac{\langle\Phi|{\tilde c}^{\dagger}_{{\bm k},\sigma}{\tilde c}_{{\bm k},\sigma}|\Phi\rangle}{\langle\Phi|\Phi\rangle}, 
\end{equation}
which can be calculated as $n({\bm k})=1-n/2-W({\bm k})$ owing to the sum rule \cite{Stephan_nk}. 
It has been established that $n({\bm k})$ of the Gutzwiller wavefunction exhibits a discontinuity at ${\bm k}_{\rm F}$ in the metallic phase 
[Figs. \ref{fig:EkWk}(g), \ref{fig:EkWk}(h), \ref{fig:EkWk}(j), and \ref{fig:EkWk}(k)] \cite{GutzwillerWF,MetznerPRL,MetznerPRB,YokoyamaShiba1dGW,Gros1dGW}. 
The value of this discontinuity is called the quasiparticle weight, which is represented by $Z$ in this paper [Figs. \ref{fig:ZWe}(c), \ref{fig:ZWe}(f), \ref{fig:ZWe}(i), and \ref{fig:ZWe}(l)]. 
The volume inside the Fermi surface of the Gutzwiller wavefunction is the same as the noninteracting Fermi sea (cf. Luttinger's theorem \cite{LuttingerTheorem}). 
The same volume as the noninteracting Fermi sea and $Z\ne 0$ in the metallic phase are usually identified as evidence of a Fermi liquid. 
In this sense, the Gutzwiller wavefunction can be regarded as a Fermi-liquid-like state. 
\par
As illustrated in Figs. \ref{fig:ZWe}(c) and \ref{fig:ZWe}(f), the quasiparticle weight $Z$ on the chain and plane decreases continuously to zero toward the Mott transition. 
The Brinkman-Rice picture is based on this behavior: $Z\rightarrow 0$ implies the divergence of the effective mass $m^*$, because $m^*\propto 1/Z$ in the Fermi-liquid theory, 
assuming that the renormalization of $m^*$ is only due to the $\omega$ dependence of the self-energy \cite{BrinkmanRice}. 
If electronic excitation can essentially be represented by the single mode of the Fermi-liquid quasiparticle with $m^*\rightarrow\infty$, 
the Mott transition should be characterized by the flattening of the dispersion relation toward the Mott transition, 
as is widely believed according to the Brinkman-Rice picture. 
\par
In contrast, as discussed in Secs. \ref{sec:spectralfunction}--\ref{sec:spectralweight}, the results on the chain and plane 
indicate that the mode crossing the Fermi level, which is continuously deformed from a noninteracting band at $n=0$, 
does not become flat toward the Mott transition, but loses its spectral weight for $\omega>0$, 
even if the ground state is assumed to be a Fermi-liquid-like state with the same volume as the noninteracting Fermi sea 
and a nonzero $Z$ that decreases continuously to zero toward the Mott transition. 
\par
The results exhibiting $Z\ne 0$ on the chain [Figs. \ref{fig:EkWk}(g) and \ref{fig:ZWe}(c)] are due to the Gutzwiller wavefunction \cite{MetznerPRL,MetznerPRB}. 
In a 1D system, the low-energy properties are generally described as a Tomonaga-Luttinger liquid \cite{HaldaneTLL,TomonagaTLL,LuttingerTLL,MattisLiebTLL} 
where $Z=0$ \cite{GiamarchiBook,EsslerBook}. 
Thus, the picture of $m^*\rightarrow\infty$ for the Fermi-liquid quasiparticle is generally inapplicable to 1D systems. 
Nevertheless, the gradual loss of the spectral weight from the dispersing mode that exhibits the momentum-shifted magnetic dispersion relation 
in the small-doping limit has been shown in 1D systems \cite{Kohno1DHub,Kohno1DtJ} 
as well as in 2D systems \cite{Kohno2DHub,Kohno2DtJ}. 
This implies that this characteristic is general and fundamental in the Mott transition, 
regardless of whether the ground state is like a Fermi liquid or not. That is, 
this characteristic is not highly sensitive to the ground-state properties or dimensionality, but would generally be robust in the Mott transition. 
\par
For the ladder and bilayer, $Z$ remains nonzero even in the limit of $n\rightarrow 1$ [Figs. \ref{fig:EkWk}(h), \ref{fig:EkWk}(k), \ref{fig:ZWe}(i), and \ref{fig:ZWe}(l)], 
as in the case of the transition from a metal to a band insulator. 
Nevertheless, similarly to the cases of the chain and plane, the dispersion relation of the antibonding band is deformed 
into the momentum-shifted magnetic dispersion relation in the small-doping limit 
[Figs. \ref{fig:EkWk}(c), \ref{fig:EkWk}(f), \ref{fig:ZWe}(g), and \ref{fig:ZWe}(j)]. 
Furthermore, the spectral weight at $k_{\perp}=\pi$ decreases continuously to zero toward the Mott transition 
[Figs. \ref{fig:EkWk}(i), \ref{fig:EkWk}(l), \ref{fig:ZWe}(h), and \ref{fig:ZWe}(k)] \cite{KohnoDIS,KohnoHubLadder}, 
contrary to the conventional band picture. These results also support the general and fundamental characteristic of the Mott transition. 
\subsection{Model with Gutzwiller-wavefunction ground state} 
The characteristic discussed in this paper can also be demonstrated using a model whose ground state is the Gutzwiller wavefunction. 
It is known that the ground state of the 1D supersymmetric $t$-$J$ model with $1/r^2$ interaction ($J/t=2$) is the Gutzwiller wavefunction \cite{SUSYGS}. 
The electron-addition spectral function $A^+(k_x,\omega)$ of this model has been obtained analytically \cite{SUSYAkw}. 
According to the analytical expression of $A^+(k_x,\omega)$ \cite{SUSYAkw}, 
the dominant mode (upper edge of the continuum) that is continuously deformed from the noninteracting band at $n=0$ 
loses its spectral weight and gradually disappears toward the Mott transition. Its dispersion relation continues to disperse and becomes 
\begin{equation}
\omega=e_{\rm HS}(k_x-k_{\rm F}) 
\end{equation}
for $k_{\rm F}<k_x<2\pi-k_{\rm F}$ in the small-doping limit (Fermi momentum $k_{\rm F}\rightarrow\pi/2$), 
where $e_{\rm HS}(k_x)$ denotes the dispersion relation of the dominant mode of the spin excitation at half filling (the Haldane-Shastry model) \cite{HSSkw}: 
\begin{equation}
e_{\rm HS}(k_x)=Jk_x(\pi-k_x)/2. 
\end{equation}
This clearly demonstrates that the characteristic discussed in this paper can appear even in a system whose ground state is a Fermi-liquid-like state 
exhibiting gradual disappearance of the quasiparticle weight toward the Mott transition [Fig. \ref{fig:ZWe}(c)]. 
\subsection{Comparisons with conventional pictures} 
In conventional single-particle pictures, an electronic quasiparticle or a hole is considered as a carrier 
and the Mott transition is considered to be characterized as one of the following two possibilities \cite{ImadaRMP}: 
the divergence of the effective mass $m^*\rightarrow \infty$ or the disappearance of the carrier density $n_{\rm c}\rightarrow 0$. 
The former is based on the Fermi-liquid theory, where interaction effectively makes the electronic quasiparticle heavier. 
The latter is based on a band picture such as the mean-field approximation for the antiferromagnetic order \cite{OverhauserSDW} 
or Hubbard's decoupling approximation \cite{HubbardApprox}, where holes in a doped Mott insulator can be regarded as carriers. 
Discussions on the Mott transition have mostly focused on which picture is more appropriate and intense controversies 
have arisen, particularly in relation to cuprate high-temperature superconductors \cite{ImadaRMP,DagottoRMP}. 
For the distinction between $m^*\rightarrow \infty$ and $n_{\rm c}\rightarrow 0$, 
the ground-state properties such as the quasiparticle weight $Z$, antiferromagnetic order, and sign of the Hall coefficient are important. 
\par
In contrast, the characteristic discussed in this paper is not highly sensitive to the ground-state properties 
\cite{Kohno1DHub,Kohno2DHub,Kohno1DtJ,Kohno2DtJ,KohnoDIS,KohnoRPP,KohnoSpin,KohnoAF,KohnoHubLadder}. 
The quasiparticle weight $Z$ or the presence or absence of a spin gap or antiferromagnetic order in a Mott insulator is not significant. 
In fact, essentially the same characteristic of the Mott transition appears on the square lattice [Figs. \ref{fig:EkWk}(d), \ref{fig:EkWk}(j), \ref{fig:ZWe}(d), and \ref{fig:ZWe}(e)] 
and on the chain [Figs. \ref{fig:EkWk}(a), \ref{fig:EkWk}(g), \ref{fig:ZWe}(a), and \ref{fig:ZWe}(b)], 
although the Gutzwiller wavefunction on a square lattice exhibits an antiferromagnetic long-range order at half filling \cite{GWAForderLi,GWAForderEnt}, 
whereas that on a chain does not \cite{GWspinPRL,GWspinPRB}. 
Instead, the existence of spin excitation in the energy regime that is much lower than the charge gap in a Mott insulator is important for this characteristic. 
This spin--charge separation can be regarded as a defining factor of a Mott insulator \cite{KohnoHubLadder}. 
In fact, in a band insulator, spin--charge separation does not occur; the lowest spin- and charge-excitation energies are the same as the band gap 
because the excitations are described in terms of electronic single particles \cite{KohnoRPP,KohnoAF}. 
\par
It should be noted that the spin--charge separation in the metallic phase of a 1D system means 
that excitations in the low-energy limit are described in terms of independent spin and charge excitations \cite{GiamarchiBook,TakahashiBook,EsslerBook}, rather than electronic quasiparticles. 
The lowest excitation energies for the spin $\Delta_{\rm s}$ and charge $\Delta_{\rm c}$ of the order of $1/N_{\rm s}$ are different. 
In a Mott insulator, the spin--charge separation is more robust [$\Delta_{\rm s}\ll\Delta_{\rm c}=O(U)$ for Coulomb repulsion $U\gg t$; 
$\Delta_{\rm s}\ll\Delta_{\rm c}=\infty$ in the $t$-$J$ (Heisenberg) model] and general, regardless of the dimensionality. 
\par
Although there are physical quantities that can distinguish between an insulator and a metal, such as the Drude weight \cite{KohnDrudeWeight}, 
the characterization of the Mott transition should reflect a general characteristic of a Mott insulator that can distinguish a Mott insulator from a band insulator \cite{KohnoHubLadder}. 
The above-mentioned spin--charge separation in a Mott insulator provides such a characteristic. 
Because the characteristic of the Mott transition discussed in this paper reflects the spin--charge separation of a Mott insulator, it would be general 
regardless of the dimensionality \cite{Kohno1DHub,Kohno2DHub,Kohno1DtJ,Kohno2DtJ,KohnoDIS,KohnoRPP,KohnoSpin,KohnoAF,KohnoHubLadder}. 
Although an antiferromagnetic order may be considered as important in the Mott transition, it is not essential to the Mott transition. 
This is because not only an antiferromagnetically ordered insulator, but also a spin liquid, which is an insulator exhibiting spin excitation 
(with or without a spin gap) in the energy regime that is much lower than the charge gap without a magnetic order, is usually regarded as a Mott insulator. 
The quasiparticle weight $Z$ in the metallic phase or structural instability is not essential to the Mott transition either, 
because the Mott transition can occur even on a chain with $Z=0$ or without being accompanied by lattice distortion. 
\subsection{Doping-induced states} 
The emergence of electronic states in the Hubbard gap upon doping a Mott insulator has been recognized 
since the early 1990s \cite{Eskes,DagottoDOS}, but the interpretations thereof have been controversial. 
The emergent states have been interpreted as part of the upper Hubbard band that is quickly shifted 
by doping \cite{SakaiImadaPRL,SakaiImadaPRB}, composite-particle states \cite{ImadaCofermionPRL,ImadaCofermionPRB,PhillipsRMP,PhillipsRPP}, 
and a spin-polaron shake-off band \cite{EderOhtaIPES,EderOhta2DHub}. 
In these interpretations, the mode of the emergent states is essentially separated by an energy gap 
from the mode around the Fermi level, even if the spin excitation of a Mott insulator is gapless 
\cite{SakaiImadaPRL,SakaiImadaPRB,ImadaCofermionPRL,ImadaCofermionPRB,PhillipsRMP,PhillipsRPP,EderOhtaIPES,EderOhta2DHub}. 
In contrast, another interpretation is that the emergent states are essentially the spin-excited states 
that exhibit the magnetic dispersion relation shifted by ${\bm k}_{\rm F}$ in the electronic spectrum 
\cite{Kohno1DHub,Kohno2DHub,Kohno1DtJ,Kohno2DtJ,KohnoDIS,KohnoRPP,KohnoSpin,KohnoAF,KohnoHubLadder}. 
If the magnetic excitation of a Mott insulator is gapless, the mode of the emergent electronic states 
should also be gapless in the small-doping limit. 
\par
The behavior of the characteristic mode discussed in this paper can also be understood in the final interpretation above 
\cite{Kohno1DHub,Kohno2DHub,Kohno1DtJ,Kohno2DtJ,KohnoDIS,KohnoRPP,KohnoSpin,KohnoAF,KohnoHubLadder}, 
even if a Fermi-liquid-like ground state is assumed [Figs. \ref{fig:Akw}(b), \ref{fig:Akw}(e), \ref{fig:Akw}(h), \ref{fig:Akw}(k), \ref{fig:ZWe}(a), \ref{fig:ZWe}(d), \ref{fig:ZWe}(g), and \ref{fig:ZWe}(j); 
Eqs. (\ref{eq:ektlimit}), (\ref{eq:eskchain}), (\ref{eq:eskplane}), (\ref{eq:eskladder}), (\ref{eq:ekchainlimit}), and (\ref{eq:ekplanelimit})]. 
When viewed from the low-electron-density side, this mode is continuously deformed from a noninteracting band at $n=0$, 
gradually losing its spectral weight toward the Mott transition 
[Figs. \ref{fig:EkWk}(a), \ref{fig:EkWk}(c), \ref{fig:EkWk}(d), \ref{fig:EkWk}(f), \ref{fig:EkWk}(g), \ref{fig:EkWk}(i), \ref{fig:EkWk}(j), \ref{fig:EkWk}(l), 
\ref{fig:ZWe}(a), \ref{fig:ZWe}(b), \ref{fig:ZWe}(d), \ref{fig:ZWe}(e), \ref{fig:ZWe}(g), \ref{fig:ZWe}(h), \ref{fig:ZWe}(j), and \ref{fig:ZWe}(k)], 
which implies that this mode also has the same origin as a noninteracting band at $n=0$. 
\subsection{Physical picture of Mott transition} 
The physical picture of this characteristic of the Mott transition has been described 
as follows \cite{Kohno1DHub,Kohno2DHub,Kohno1DtJ,Kohno2DtJ,KohnoDIS,KohnoRPP,KohnoSpin,KohnoAF,KohnoHubLadder}: 
From the metallic side, the charge degrees of freedom freeze toward the Mott transition, 
while the electronic motion is preserved in the spin degrees of freedom. 
This picture has been derived based on the spectral feature indicating that an electronic mode representing an electronic particle with spin and charge 
gradually loses its identity (spectral weight) toward the Mott transition, while the dispersion relation is continuously reduced 
to the magnetic dispersion relation shifted by ${\bm k}_{\rm F}$ in the Mott transition. 
From the insulating side, spin excitation emerges as electronic excitation because the charge character is added by doping. 
This picture has been derived based on the spectral feature indicating that the electronic excitation in the small-doping limit exhibits 
the magnetic dispersion relation shifted by ${\bm k}_{\rm F}$ 
(a spin-excited state at ${\bm k}={\bm p}$ from the ground state with $N_{\rm e}=m$ can overlap 
with an electron-addition excited state at ${\bm k}={\bm p}+{\bm k}_{\rm F}$ from the ground state with $N_{\rm e}=m-1$ 
where an electron with ${\bm k}={\bm k}_{\rm F}$ on the Fermi surface is removed \cite{KohnoDIS}). 
The emergence in the electronic spectrum implies that the excitation not only has a spin character, but also gains a charge character owing to doping 
(the electronic excitation should have the same quantum numbers as an electron). 
\par
The above picture contrasts with conventional single-particle pictures: 
the electronic quasiparticle becomes extremely heavy and immobile when $m^*\rightarrow\infty$, 
and the number of mobile holes disappears (full filling of electrons) when $n_{\rm c}\rightarrow 0$. 
In these single-particle pictures, the decoupling of the spin and charge degrees of freedom toward the Mott transition is not considered. 
These pictures do not explain how a metallic state changes into a Mott insulating state 
that exhibits spin excitation in the energy regime that is much lower than the charge gap (the spin--charge separation characteristic of a Mott insulator). 
\section{Summary} 
Electron-addition excitation from the Gutzwiller wavefunction was investigated in the 1D, 2D, ladder, and bilayer $t$-$J$ models 
in the single-mode approximation using a Monte Carlo method. 
In all of these models, the numerical results demonstrated that an electronic mode that is continuously deformed from a noninteracting band at zero electron density loses its spectral weight 
and gradually disappears toward the Mott transition, exhibiting essentially the magnetic dispersion relation shifted by the Fermi momentum in the small-doping limit. 
Thus, this characteristic would be general and fundamental in the Mott transition, 
regardless of the dimensionally, lattice structure, and even the presence of a spin gap or antiferromagnetic long-range order in a Mott insulator. 
Because this characteristic can be obtained simply by assuming that the ground state is the Gutzwiller wavefunction, 
it would not depend on the ground-state details, but rather, reflects a general characteristic of a Mott insulator, 
namely, spin--charge separation (the existence of spin excitation in the energy regime that is much lower than the charge gap). 
\par
This result contrasts with the conventional single-particle pictures such as the Fermi-liquid quasiparticle picture and band picture (mean-field approximation). 
In these pictures, the divergence of the effective mass (the flattening of the dispersion relation) or disappearance of the carrier density 
is considered as the essence of the Mott transition, 
where the spectral-weight loss from a dispersing mode or continuous evolution to the spin excitation of a Mott insulator is not expected. 
Meanwhile, the characteristic mode shown in this paper has the same origin not only as a noninteracting band at zero electron density, 
but also as spin-excited states in a Mott insulator, and is continuously deformed between these two limits, even under the assumption of a Fermi-liquid-like ground state. 
\par
In the future, experimental confirmation of this characteristic, as well as a reexamination of the material properties near the Mott transition 
that have been interpreted in conventional single-particle pictures, will be useful for deeper understanding of the Mott transition and electronic states in strongly correlated systems. 
\begin{acknowledgments} 
The author would like to thank S. Uji and R. Kaneko for helpful discussions. 
This work was supported by JSPS KAKENHI (Grant No. JP26400372) and the JST-Mirai Program (Grant No. JPMJMI18A3), Japan. 
The numerical calculations were partly performed on the supercomputer at the National Institute for Materials Science. 
\end{acknowledgments}

\end{document}